# Experimental Evaluation of Multi-Hop Cellular Networks using Mobile Relays

J. Gozalvez and B. Coll-Perales
UWICORE, Ubiquitous Wireless Communications Research Laboratory, http://www.uwicore.umh.es
Miguel Hernandez University of Elche, Avda. de la Universidad, s/n, 03202 Elche, Spain
j.gozalvez@umh.es, bcoll@umh.es

**Abstract-** Future wireless networks are expected to provide high bandwidth multimedia services in extended areas with homogeneous Quality of Service (QoS) levels. Conventional cellular architectures might not be able to satisfy these requirements due to the effect of surrounding obstacles and the signal attenuation with distance, in particular under Non Line of Sight (NLOS) propagation conditions. Recent studies have investigated the potential of Multi-hop Cellular Networks (MCNs) to overcome the traditional cellular architecture limitations through the integration of cellular and ad-hoc networking technologies. However, these studies are generally analytical or simulation-based. On the other hand, this paper reports the first experimental field tests that validate and quantify the benefits that MCNs using mobile relays can provide over traditional cellular systems.

## 1. INTRODUCTION

Cellular systems have significantly evolved over the past decades with the emergence of new and efficient radio access technologies, and the implementation of advanced techniques to increase the network capacity and support higher data rates. The cellular architecture has also evolved with the introduction of 3.5G and 4G standards, although the evolution has been generally confined to the traditional infrastructure-centric architecture where each Mobile Station (MS) directly communicates with the Base Station (BS). Single-hop cellular architectures currently fail to efficiently provide homogeneous QoS levels across the cell coverage area, mainly due to the effect of surrounding obstacles and the signal attenuation with the distance that result in lower QoS levels at cell boundaries or shadow areas. These limitations can be partially overcome by increasing the number of BSs and improving the spatial reuse of the scarce spectrum resources, although this approach can have a significant economic cost and face social rejection against the deployment of additional BSs. An important novelty was introduced by the LTE-Advanced standard through the integration of relaying techniques into cellular systems [1]. Relaying techniques are expected to significantly improve the system capacity and user-perceived QoS through the substitution of long-distance, and generally NLOS, cellular links by various multi-hop links with improved link budgets [1]. Although there are still significant challenges to be solved, the integration of relaying techniques into cellular systems represents a significant step towards overcoming the fundamental communication and radio propagation limits of traditional single-hop cellular architectures that make the provision of  homogenous and high QoS levels across the coverage area difficult.
The integration of relaying techniques into cellular systems, usually referred to as Multi-hop Cellular Networks (MCNs), can be accomplished using either Fixed Relays (MCN-FR) or Mobile Relays (MCN-MR). Initial activities in LTE-Advanced have mainly focused on fixed and strategically located relaying stations [2]. On the other hand, MCN-MR networks use MSs to support the ad-hoc relaying capabilities in a collaborative operational framework. As a result, MCN-MR networks are characterized by a lower deployment cost but a higher management complexity. However, exploiting the communications and computing capabilities of mobile terminals offers new connectivity opportunities and future perspectives for MCN-MR networks [3]. First studies have investigated the benefits that MCN-MR networks can provide over traditional cellular architectures in terms of capacity, cell coverage, network scalability, infrastructure deployment cost, power consumption and energy efficiency [3, 4, 5]. However, these studies are either analytical or based on simulations, and there is yet the need to experimentally validate the benefits of MCN-MR technologies. In this context, this paper presents the first experimental evaluation of MCN-MRs through field tests using commercial cellular networks.

## 2. mHOP: MCN-MR EXPERIMENTAL TESTBED

Different experimental studies have analysed the integration of cellular and ad-hoc networking technologies, although to date none has focused on MCN-MRs. An interesting development is the release of Qualcomm's FlashLinq solution [6] enabling mobile devices to discover automatically neighboring terminals and communicate with them without infrastructure support. FlashLinq targets device-to-device communications, and not multi-hop cellular communications. The first experimental evaluation of MCN technologies is reported in [2], where the authors present a prototype based on the LTE-Advanced standard and using fixed relays. The conducted experiments demonstrate that MCN-FR systems can significantly improve the coverage and capacity of cellular systems. The study also confirms the need to strategically locate the fixed relay nodes at the cell edge under LOS conditions to the BS in order to maximise capacity and end-user QoS.
Despite the increased interest in MCN technologies, to the authors' knowledge, there has not yet been an experimental evaluation of MCN-MR systems. In this context, the UWICORE Laboratory at the Miguel Hernandez University of Elche (Spain) has implemented mHOP, the first MCN testbed based on mobile relays[1]. mHOP was implemented to investigate the benefits of MCN-MR over traditional cellular systems, and the conditions under which such benefits can be obtained. As a result, the platform includes an MCN-MR link and a conventional single-hop cellular link to allow for their performance comparison using software monitoring tools. Figure 1 illustrates the architecture of the mHOP testbed which currently focuses on downlink transmissions. The figure shows that the destination node can be reached from the BS through the traditional single-hop cellular link, or through the MCN-MR link. The MCN-MR link requires access to the cellular infrastructure through a hybrid Mobile Node (MN) capable of forwarding cellular data in real-time to the destination node. In particular, the hybrid MN forwards the data to the destination node using other mobile ad-hoc nodes as relay nodes. The mHOP MNs include the necessary software tools to monitor the operation and QoS of the multi-hop forwarding process.

[1] The testbed here described is an evolution of the platfom initially presented in [7].

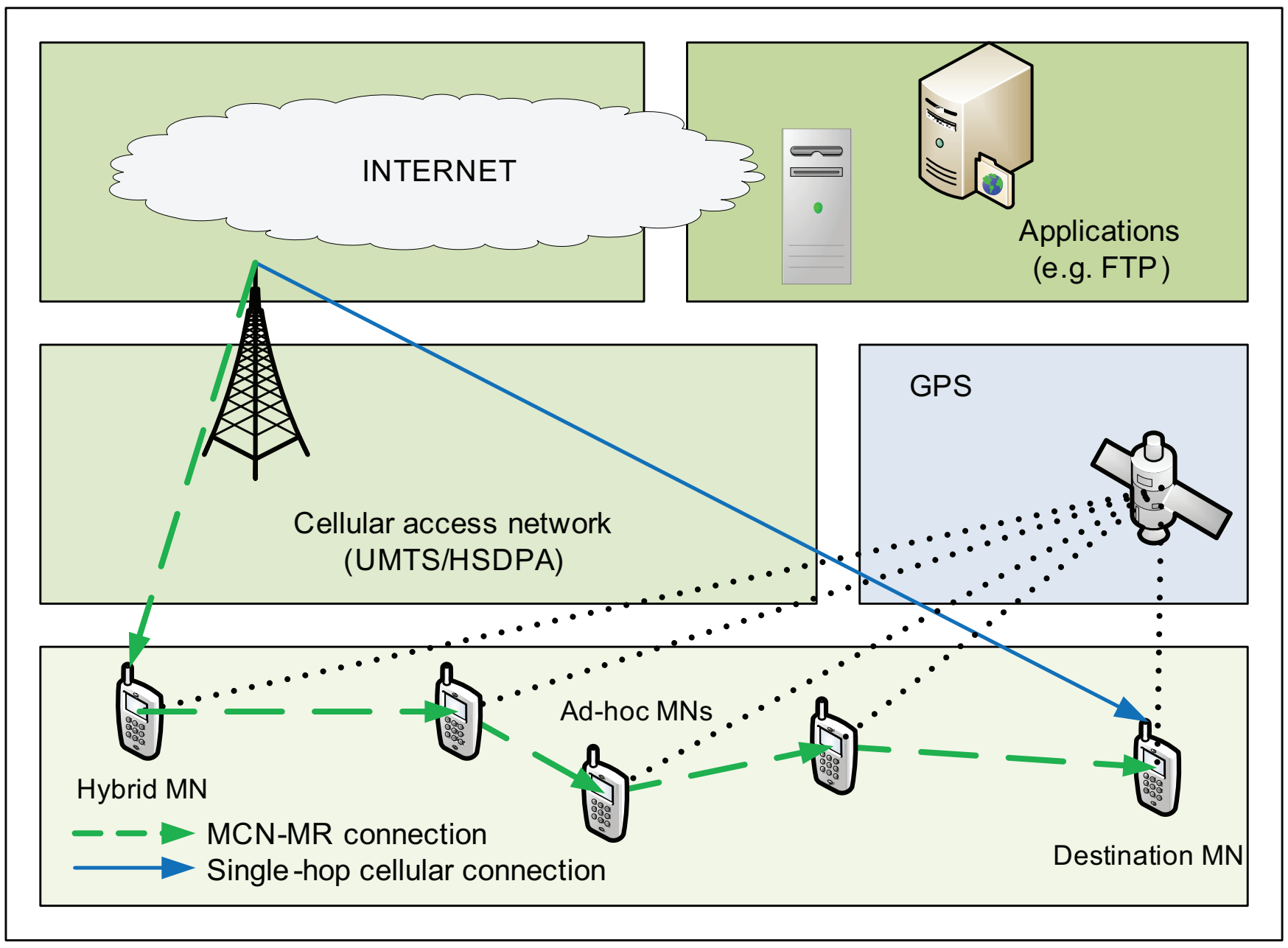

Figure 1. mHOP platform

## 2.1. Cellular Connectivity

The mHOP platform includes cellular connectivity into the hybrid MN and the destination MN using a Nokia 6720c handset. The Symbian-based terminal supports GSM/GPRS and UMTS/HSDPA, and incorporates the Nemo Handy application (http://www.anite.com). Nemo Handy is a professional network testing tool used to extensively monitor in real-time the operation of cellular networks. It provides a large set of parameters and measurement data captured over voice and video calls, FTP/HTTP data transfers, iPerf testing, HTML/WAP browsing, video streaming, SMS/MMS messaging, e-mail and ping services. Nemo Handy also allows storing the monitored parameters for post-processing using Nemo Outdoor. Nemo Outdoor offers a valuable set of key performance indicators, e.g. throughput, BLER (Block Error Rate) or RSSI (Received Signal Strength Indication), that have been very valuable to analyze the cellular QoS in mHOP. Another important feature of Nemo Handy is the possibility to lock the cellular connection to a specific radio access technology and BS. This feature facilitated a more stable testing environment in order to compare single-hop cellular and MCN-MR connections. Finally, Nemo Handy also provides spatial and time synchronization through an external GPS connected via Bluetooth. The GPS data has been used to track the MS position and geo-reference all the performance measurements.

## 2.2. Ad-Hoc Connectivity

mHOP ad-hoc MNs consist of laptops transmitting over IEEE 802.11. In addition to their built-in wireless interface, the laptops have been equipped with a Wireless ExpressCard integrating the AR9280 Atheros chipset. The laptops use their Wireless ExpressCard for ad-hoc transmissions, and they monitor the transmitted and received packets using both their built-in wireless interface and a virtual interface created using the Wireless ExpressCard. The laptops have also been equipped with an USB GPS receiver that, together with the Network Time Protocol (NTP), allows time-synchronizing all mHOP devices. In addition, the GPS is used to geo-reference all the logged radio measurements.

The ad-hoc MNs operate under Linux using the Ubuntu 10.04 distribution that includes the Linux Kernel 2.6.32. This solution was chosen due to the availability of open tools to configure parameters at the 802.11 physical and MAC layers. In particular, mHOP uses the *Compat-Wireless* and *iw* packages to configure the wireless interfaces. These packages are used to: 1) create the ad-hoc network that mHOP ad-hoc MNs will join to establish the MCN-MR end-to-end connection, 2) configure the built-in wireless interface, and 3) configure the Wireless ExpressCard's virtual interface in monitor mode. The ad-hoc MNs incorporate a packet sniffer software developed at UWICORE to capture 802.11 traffic. The sniffer uses the open source *libpcap* library provided by Linux and employed in tools such as Wireshak/T-Shark or Kismet. Having access to the raw captured packets allows a much faster and customized filtering and post-processing than using certain network testing tools. The ad-hoc MNs time- and geo-reference all the captured packets using a USB-based GPS module and a simple application developed (using the Linux *libgps* library) to collect the time, latitude and longitude with a refresh periodicity of 1Hz.

The hybrid MN is an ad-hoc MN that also acts as a gateway between the cellular and 802.11 multi-hop ad-hoc links. To this aim, the hybrid MN is also implemented on a laptop with all previously described 802.11 features, and uses a Nokia 6720c terminal as modem to provide the cellular link required in MCN-MR connections. The hybrid MNs' routing tables have been modified to allow for the real-time forwarding of information from the cellular to the multi-hop ad-hoc 802.11 links. Finally, the hybrid MN includes two GPS receivers and the cellular and ad-hoc software monitoring tools previously described.

## 3. EXPERIMENTAL MCN-MR PERFORMANCE EVALUATION

This section reports the first experimental MCN-MR field trials over commercial live cellular networks. Orange Labs in Spain supported the field trials conducted over different locations in the city of Elche (Spain). The locations were chosen to identify and evaluate the conditions under which MCN-MR technologies can overcome the limitations of traditional single-hop cellular systems.

The field trials consider downlink transmissions from a local FTP server located at the UWICORE Laboratory. The cellular links are configured to operate over HSDPA, and the ad-hoc ones over 802.11g at 2.4GHz (maximum bit rate of 54Mbps)[2]. It is important to note that the testing BSs are not barred, and therefore other users might be active. Although this configuration guarantees realistic operating conditions, it requires a careful treatment to adequately interpret the field measurements. Figure 2 depicts the QoS experienced by a single-hop cellular user as a function of the distance to its serving BS (the results are shown as relative performance since the actual measurements have been divided by the maximum value experienced for each parameter during the test). The dotted line represents the percentage of time that the user was assigned radio resources (*usage*), and the dash-dotted line represents the user's Channel Quality Indicator (CQI)[3]. Figure 2 shows that the CQI decreases with the distance to the serving BS. However, the user throughput remains relatively constant despite the degradation of the channel quality conditions. This is the case because the network detects the channel degradation and compensates it by increasing the *usage* of radio resources as the user moves away from the serving BS. This results in the user throughput level being relatively independent of the distance to the serving BS for the measurements reported in Figure 2. However, if the throughput is divided by the *usage* parameter (*normalized throughput* in Figure 2), it is then possible to observe the dependence of the cellular performance with the channel quality (and distance). To reduce the impact of the radio resource management policies on the conducted study, the *normalized throughput* has been selected as the reference parameter.

[2] The data rate and transmission power are dynamically modified based on the link quality conditions (the maximum transmission power was set to 19dBm).

The field trials have been carried out under diverse configurations combining different number of hops and hop distances. For example, a 3hops-40m configuration refers to an MCN-MR connection composed of one cellular link and two 802.11g ad-hoc links, and a distance between MNs of 40 meters. If the tests involve mobility, the MNs approximately maintain the distance during the route. The tests start when the destination node launches a script that executes a file download through the MCN-MR link. The MCN-MR link requires that the hybrid MN transforms the transport blocks received through the cellular link to 802.11g packet data units (MAC PDU); the MAC PDU size has been set to 1564 bytes. As a result, the hybrid MN stores the cellular transport blocks in a buffer until an 802.11g MAC PDU is filled. Once the 802.11g packet is formed, the hybrid MN transmits the MAC PDU to the next ad-hoc MN using its 802.11g wireless interface. Ad-hoc MNs located between the hybrid MN and the destination MN act as relays with the multi-hop path being predefined at the start of each field trial. Since HSDPA is characterized by lower data rates than 802.11g, the MCN-MR connection is upper-bounded in performance by its cellular link. For a fair comparison with the single-hop cellular link, and following the observations reported from Figure 2, the MCN-MR throughput is here reported as the ratio between the MCN-MR user throughput measured at the destination MN and the usage of the cellular link experienced by the hybrid MN.

[3] The HSDPA standard uses CQI estimates to provide an indication of the channel quality and recommend transmission modes (transport block size, number of parallel codes, modulation scheme, etc.). High CQI values indicate good channel quality conditions.

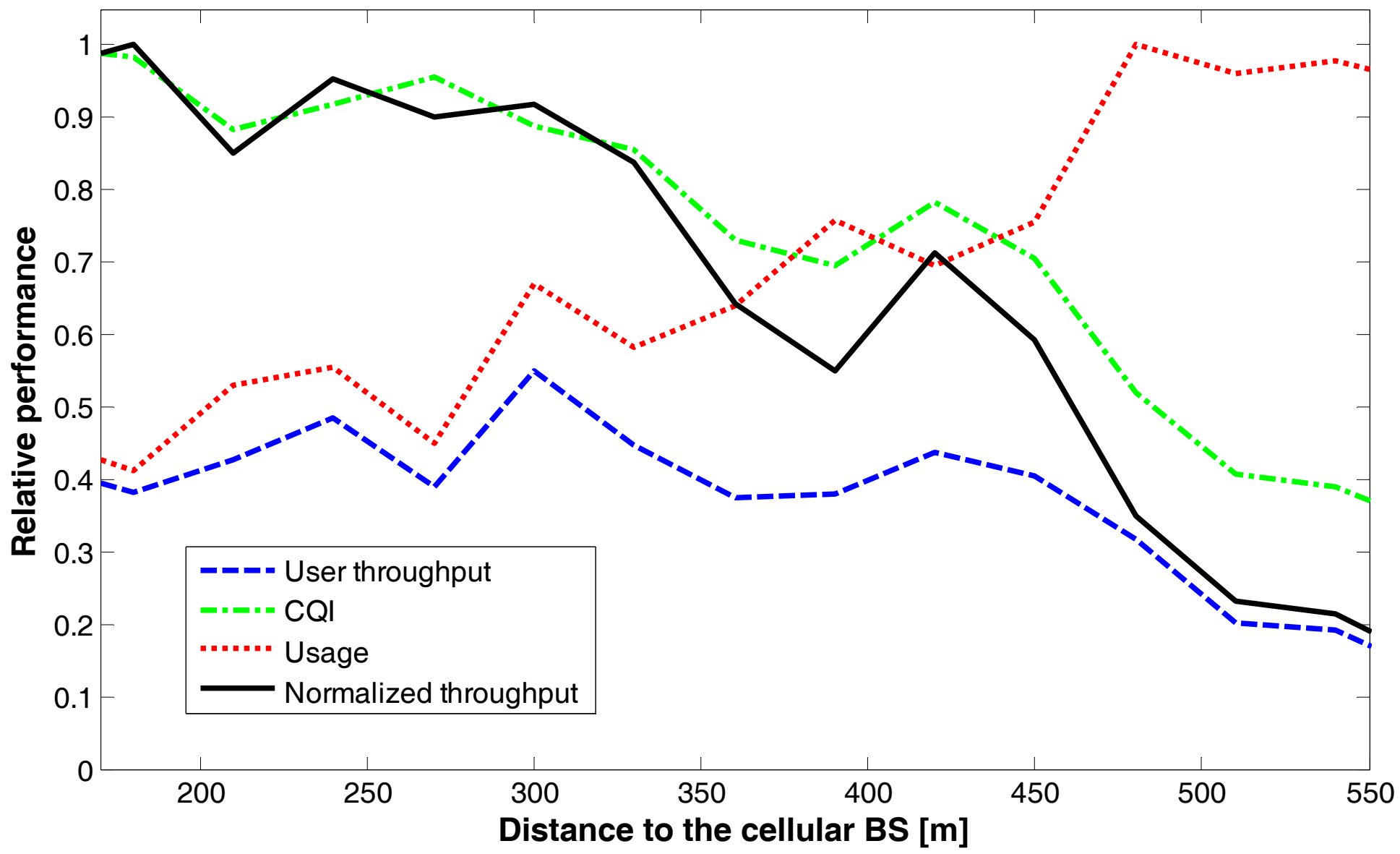

Figure 2. Cellular performance as a function of the distance to the serving BS

### 3.1. QoS at Large Distances to the Serving BS

A major motivation behind the development of relaying techniques in cellular systems is the possibility to provide the same QoS experienced by users close to a BS to users at larger distances across the cell area. This can be done by substituting long-distance and NLOS single-hop cellular links by MCN connections using either fixed or mobile relays. To extend the QoS experienced close to the BS to larger distances from the serving BS, multi-hop MCN links need to experience good LOS conditions and/or shorter distances among the communicating nodes.

Figure 3 represents the *normalized throughput* experienced by a single-hop cellular link and MCN-MR links under various configurations in a typical urban setting. The solid line in Figure 3 represents the *normalized throughput* experienced by a single-hop cellular user walking away from the BS. As previously discussed, the cellular QoS decreases with the distance as a result of the higher signal losses and lower link quality. The measurements reported in Figure 3 have been obtained using Nemo Handy, which evaluates the cellular QoS every 0.2 seconds. However, the reported measurements are averaged over time periods of one second given the GPS updating rate (1Hz). The cellular HSDPA link can reach data rates above 10 Mbps using various channelization codes (to a maximum of 15) and the 16QAM modulation. However, its performance is very dependent on the link quality conditions, and thereby on the distance between the cellular user and its serving BS. The cellular link used the high level modulation mode (16QAM) and an average of 10 channelization codes close to the BS. As the user walked away from the BS and the channel conditions degraded, the cellular link was characterized by a more robust modulation scheme (QPSK) and a lower number of channelization codes.

Figure 3 also depicts the *normalized throughput* measured at a destination node under three different MCN-MR configurations consisting of 2, 3 and 4 hops with increasing distances between ad-hoc MNs from 60 to 120 meters. The location of the destination node was fixed at 780 meters away from the BS. On the other hand, the location of the hybrid MN and ad-hoc MNs vary depending on the MCN-MR configuration. It is important to note that the higher the number of hops, the closer will be the hybrid MN to the BS, and thereby the higher its cellular QoS. The links between ad-hoc MNs are characterized by LOS conditions with pedestrians continuously crossing over. As previously explained, the measured MCN-MR performance is generally upper-bounded by the cellular QoS given the lower HSDPA data rates compared to 802.11g (this was actually the case for the measurements reported in Figure 3). As a result, the destination node achieved the same *normalized throughput* as the hybrid MN for all the analysed MCN-MR configurations. The *normalized throughput* at the destination node is therefore depicted at the position at which the hybrid MN was located. For example, the destination node located 780 meters away from the BS achieved a *normalized throughput* of approximately 4.2Mbps when operating an MCN-MR link with 4 hops and distances among ad-hoc MNs of 120 meters. This performance is reduced to 3.5Mbps if the distance among ad-hoc MNs decreases to 90 meters since the hybrid MN is further away from the BS compared to the case in which the distance between ad-hoc MNs is equal to 120 meters. These experimental results clearly show that MCN-MR technologies can provide nodes located at large distances from the serving BS with high QoS levels close to those perceived by nodes located close to the serving BS. This will in turn result in a more efficient use of the radio resources.

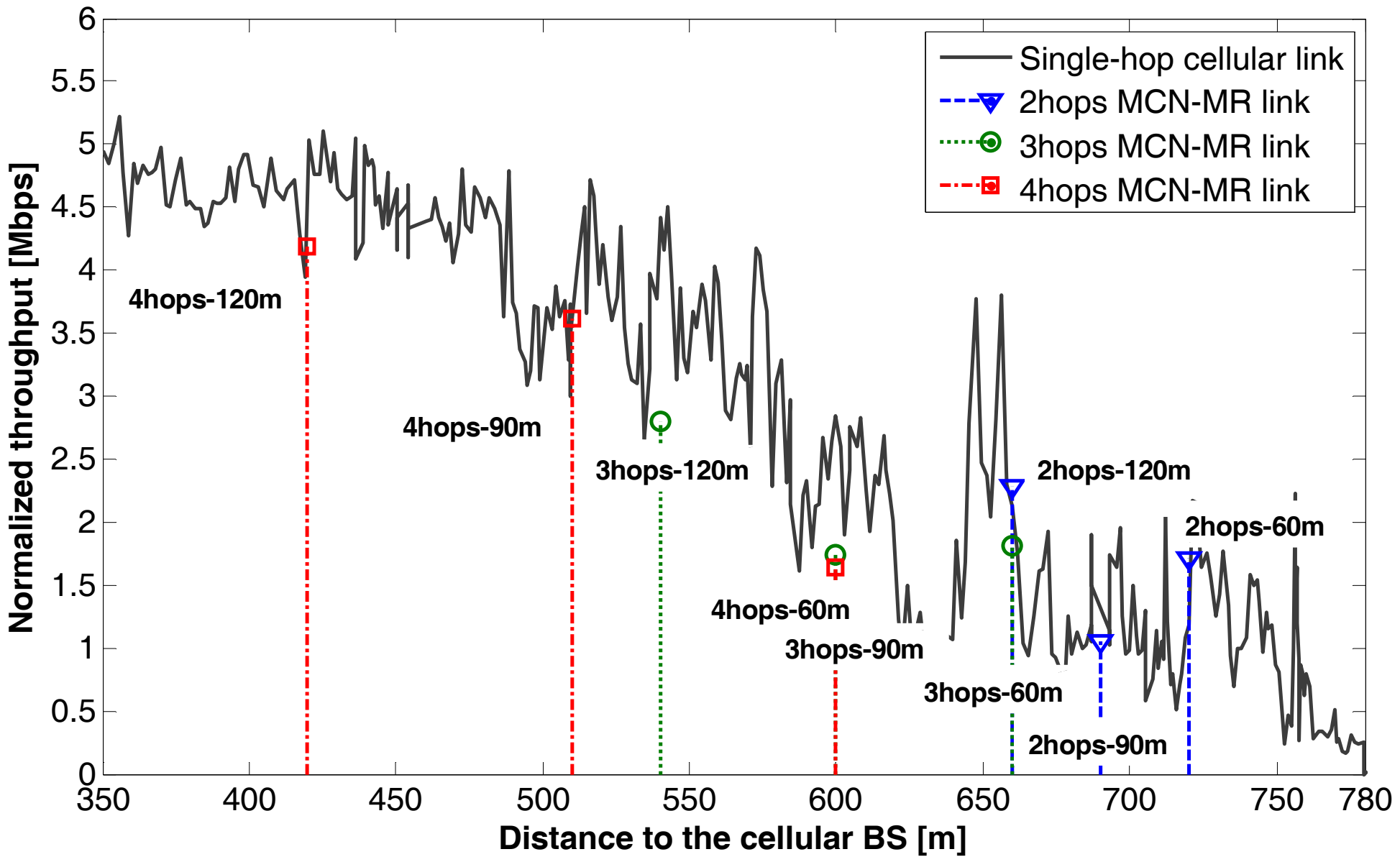

Figure 3. MCN-MR capacity to improve QoS at large distances from the serving BS

### 3.2. Coverage Extension

MCN-MR networks allow distant nodes to connect to a cellular BS through other relaying mobile nodes. This feature can facilitate extending the radio cell coverage area compared to single-hop cellular networks, which can be attractive to provide service in shadow areas or disaster areas where part of the infrastructure is temporarily unavailable. The MCN-MR cell extension capability could also be benificial to balance traffic among neighbouring BSs under traffic congestion scenarios. For example, a node located under a congested cell could connect to the cellular network using a hybrid mobile relay node located at the cell edge of a less congested neighbouring cell.

To demonstrate the capacity of MCN-MR technologies to extend the coverage area, initial tests were conducted using the Nemo Handy tool to measure the actual radius of a typical urban cell in the city of Elche. To this aim, the testing MS was forced to connect to the BS under study using Nemo Handy. Figure 4 shows the *normalized throughput* as a function of the distance to the BS. The reported measurements show that the cell radius for the conventional HSDPA single-hop cellular link is equal to approximately 650 meters. Similarly to the previous section, several MCN-MR field trials with varying configurations were conducted to verify the capacity to extend the cell radius. In particular, Figure 4 shows the MCN-MR *normalized throughput* at the mobile destination node when configured with 3 and 5 hops, and a distance between MNs of 60 and 85 meters respectively. The hybrid and ad-hoc MNs were all located on the same street with the hybrid MN being the closest to the BS and the destination node the farthest[4]. In the case of an MCN-MR link with 3 hops, the hybrid MN was at a distance of 400 meters to the BS at the start of the trial. This distance decreased to 180 meters in the case of an MCN-MR link with 5 hops. When the destination MN executes the file download, all the MNs move in the direction of the cell boundary to increase the distance to the BS. The download continues until the hybrid MN reaches the cell boundary (cellular coverage is lost). At this point, the destination node is located 120 meters and 340 meters beyond the cell limit (approximately 650 meters) in the case of the 3 hop and 5 hop MCN-MR links respectively, thereby demonstrating the capacity of MCN-MR to extend the cell radius compared to conventional single-hop cellular networks (see Figure 4). In fact, the results depicted in Figure 4 not only highlight the cell extension capability, but also the acceptable QoS levels that MCN-MR can provide over the extended coverage areas. As previously discussed, the lower cellular performance typically results in the QoS level experienced by the destination node being similar to that experienced by the hybrid MN. The hybrid MN is closer to the BS in the case of the 5 hop MCN-MR link than in the case of the 3 hop one. As a result, the 5 hop MCN-MR configuration provides higher QoS levels than the conventional single-hop cellular link and the 3 hop MCN-MR link.

[4] MCN-MR performance exhibits larger variations than conventional cellular links. This is due to the fact that the mobile nodes had to cross several streets during the experiments resulting in temporary loss of the LOS conditions.

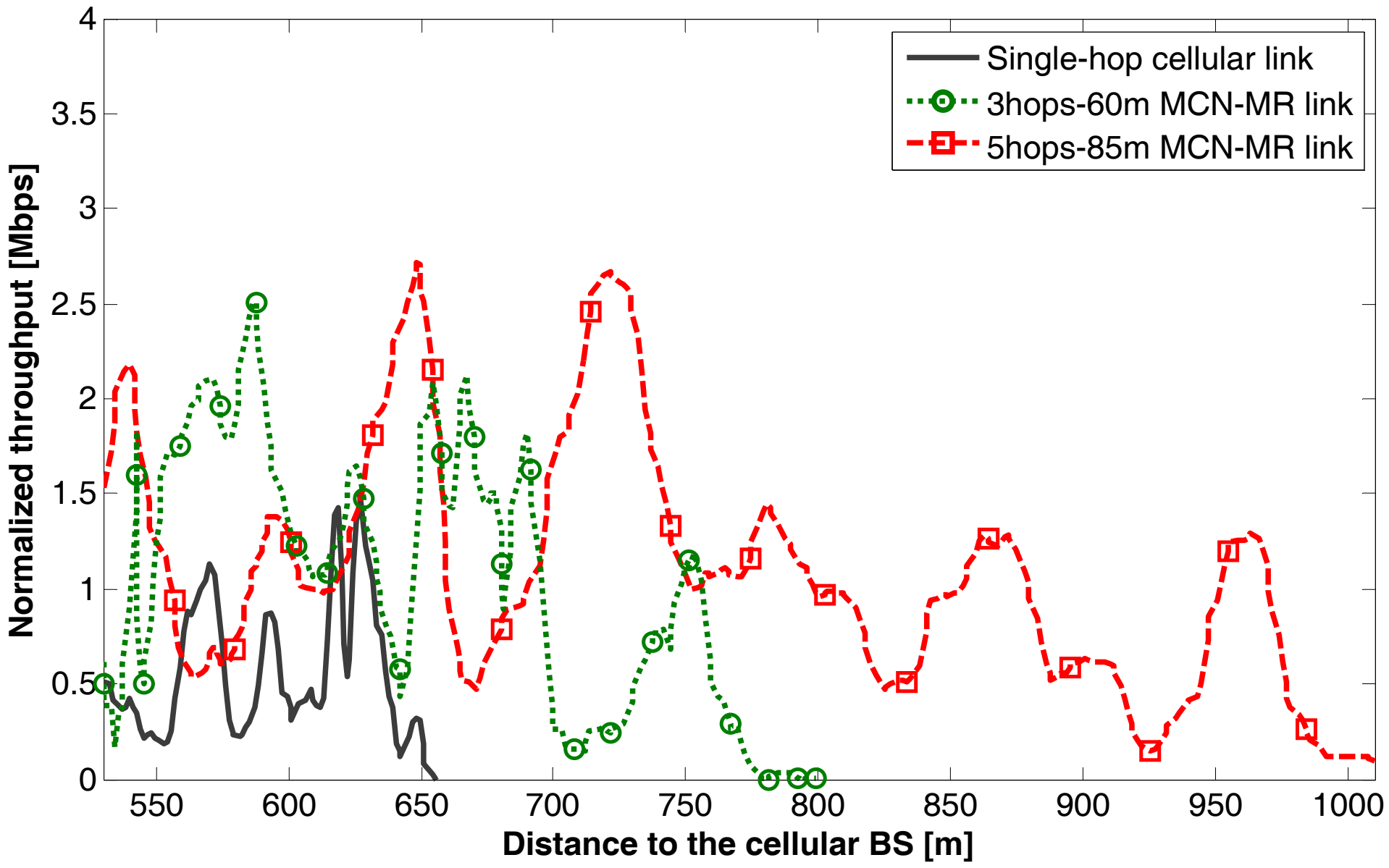

Figure 4. MCN-MR capacity to extend the radio coverage

### 3.3. Indoor QoS

Cellular trasnmissions can experience a notable QoS degradation when entering indoor environments without interior pico or femto cells. An alternative to deploying indoor BSs is the use of MCN-MR connections to extend outdoor QoS levels to indoor environments through the use of mobile relaying nodes. Field trials have been conducted in a shopping center of Elche to demonstrate this MCN-MR capacity. The outdoor cellular BS used for the experiments was located at 500 meters to the entrance of the shopping center under LOS conditions. Figure 5 shows the *normalized throughput* experienced by a conventional single-hop HSDPA connection as the MS walks inside the shopping center and is locked to the outdoor BS. The conducted experiments show that the outdoor QoS is maintained next to the entrance, but rapidly decreases as the user walks inside the shopping center due to the signal attenuation.

Three different MCN-MR configurations with 2, 3 and 4 hops, and a distance between MNs of 75 meters were also tested. At the start of the tests, the hybrid MN is located outside the shopping center (75 meters away from the entrance), while the ad-hoc MNs and the destination node are located inside. The distance of the destination node to the entrance of the shopping center increases with the number of ad-hoc hops in the MCN-MR configuration. When the destination node executes the file download, all the MNs move 75 meters inward (the hybrid MN ends up just inside the shopping center). Indoor markers using Nemo Handy were necessary to measure distances due to the poor indoor GPS signal. At the time of the experiments, the shopping center was crowded and people were continuously blocking the LOS conditions between the ad-hoc MNs. However, the results depicted in Figure 5 show that MCN-MR technologies are capable of increasing indoor QoS performance to levels close to that experienced in outdoor conditions with better cellular signal quality. The capacity to extend the cellular radio coverage is also observed in the results depicted in Figure 5 since MCN-MR is capable of maintaining active communication links at higher distances than single-hop cellular systems.

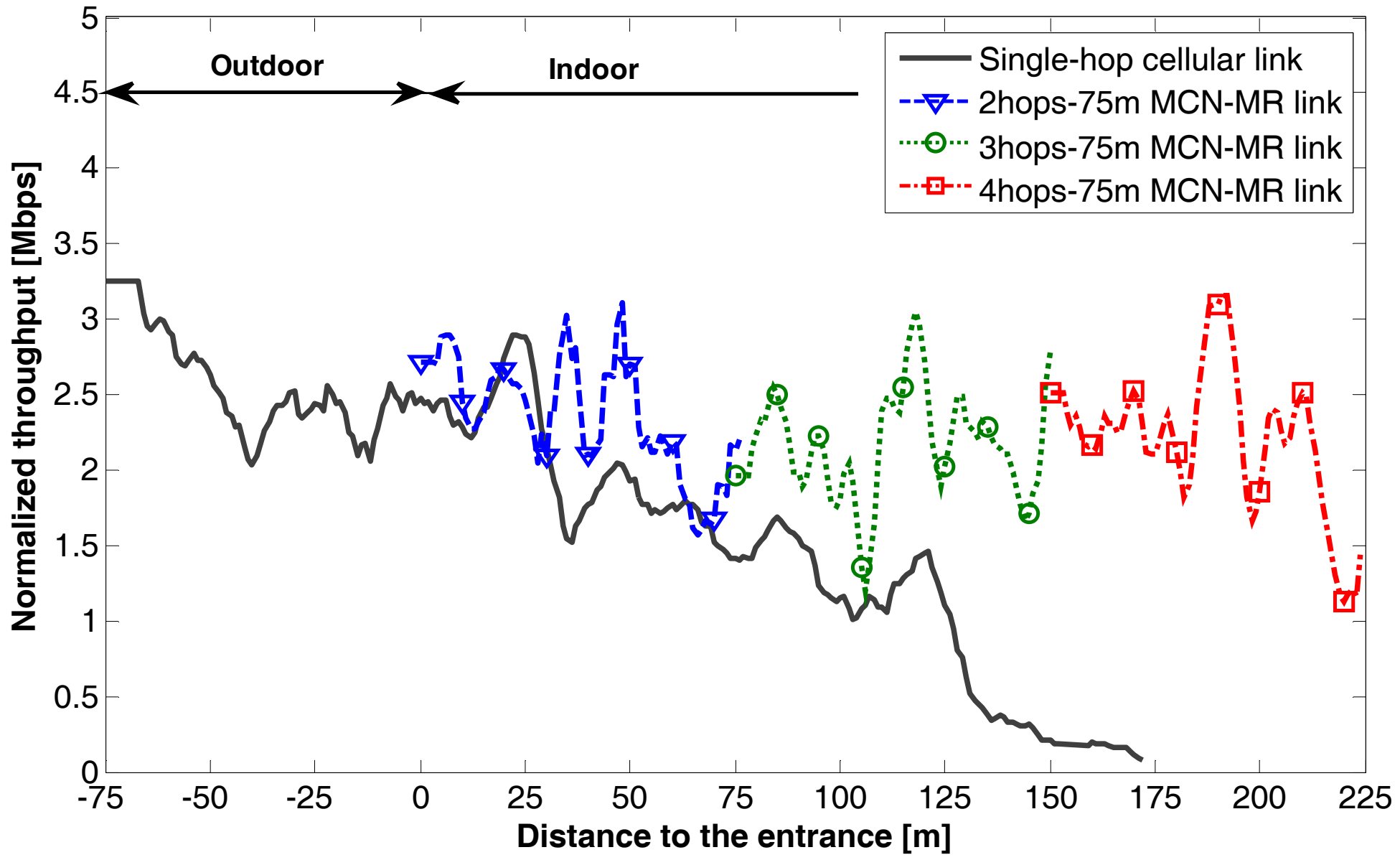

Figure 5. MCN-MR capacity to improve indoor QoS levels

**3.4. QoS under NLOS Conditions**

The presence of obstacles significantly attenuates wireless signals and can result in shadow areas, which can prevent providing homogeneous QoS levels across a cell. The field experiments reported in [2] for MCN networks with fixed relays showed that the use of relay nodes can help reduce the disadvantages experienced under single-hop cellular NLOS conditions. To do so, MCN technologies should replace large-distance NLOS single-hop cellular links by shorter-distance LOS multi-hop communications. In this context, the aim of this section is to complement the conclusions reported in [2], and demonstrate that MCN-MR networks can also enhance the single-hop cellular performance when a MS operates under NLOS conditions.

The field trials have been conducted in a typical urban cell in the city of Elche with the destination node turning around an intersection corner. This setting allows reproducing the traditional signal attenuation experienced when passing from LOS to NLOS conditions. To do so, the MS is locked to the same BS providing service under LOS conditions, i.e. a cellular handover is not allowed when turning around the corner. Figure 6 shows the *normalized throughput* experienced by a single-hop cellular link when experiencing LOS and NLOS conditions as the MS turns around the intersection corner. The negative distances represent the distance to the corner under LOS conditions, and the positive ones the NLOS distances after turning the corner. The reported measurements show how the single-hop cellular QoS rapidly decreases after turning around the corner and the MS enters NLOS conditions with the serving BS. The MCN-MR performance has been analysed under two different multi-hop configurations. Both of them consider a hybrid MN, two ad-hoc relaying MNs, and a destination node, but with different distances among the mobile nodes. At the start of the experiments, the hybrid, ad-hoc and destination MNs are placed linearly under LOS conditions with the serving BS. Once the file download is executed, the MNs walk towards the intersection corner, with the destination node being the first to turn around the corner. As the file download continues, ad-hoc MNs and finally the hybrid MN reach the corner. The point at which the hybrid MN reaches the corner is marked in Figure 6 for each MCN-MR configuration. Figure 6 also represents the *normalized throughput* experienced at the destination MN for the two MCN-MR configurations. The reported measurements show that MCN-MR technologies can improve the NLOS QoS performance to levels experienced under LOS when adequately selecting the relay nodes (the mechanisms and criteria to select relay nodes are out of the scope of this paper). The performance difference between both MCN-MR configurations is a result of the varying distance among ad-hoc nodes. A higher distance allows maintaining the connection during larger distances after turning the intersection corner. However, it also increases the time during which an 802.11g ad-hoc link experiences NLOS conditions when one of the ad-hoc MNs turns the corner. This results in temporary deeper QoS degradations observed for the configuration with 50 meters between ad-hoc MNs (Figure 6 around 40m).

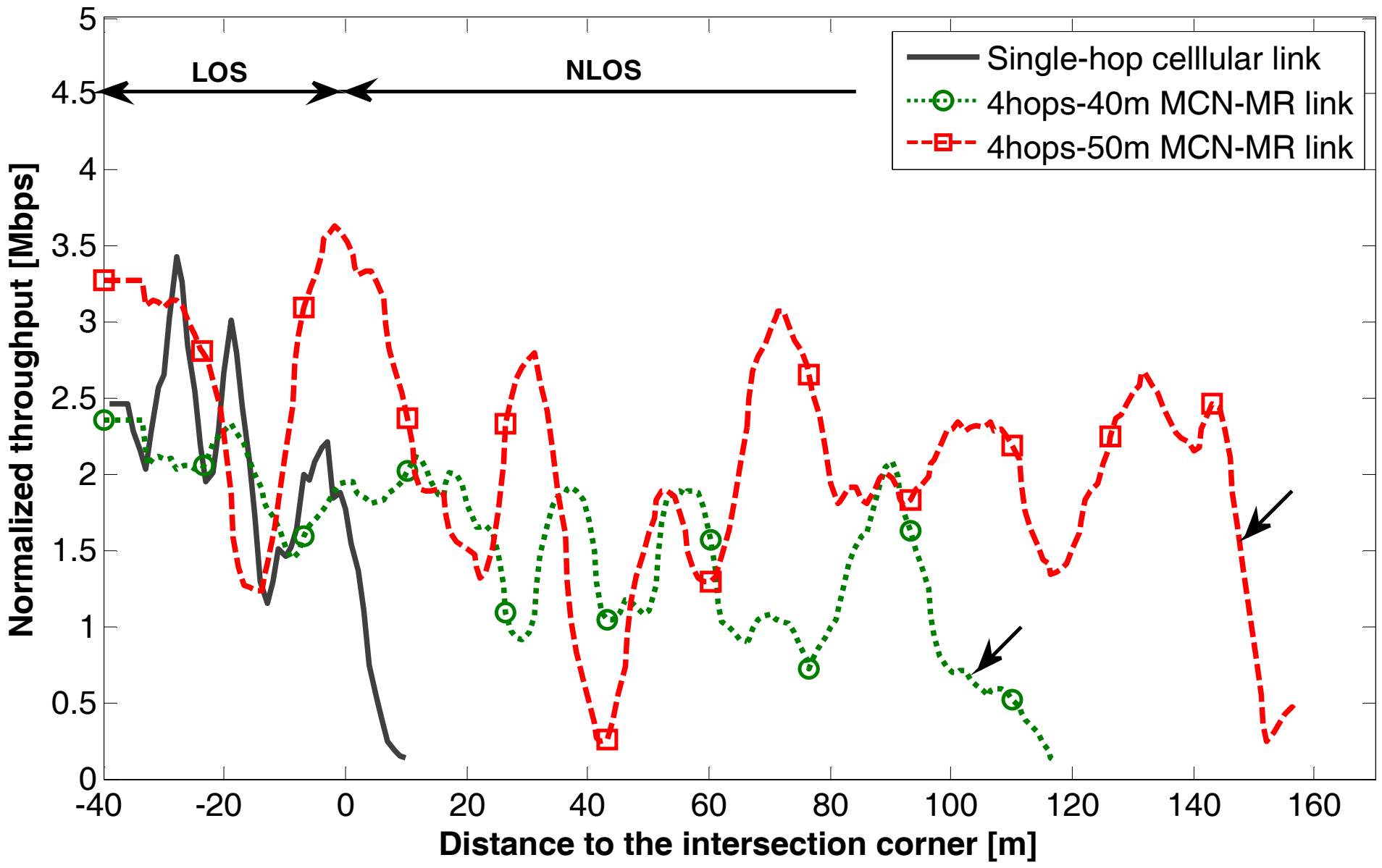

Figure 6. MCN-MR capacity to improve NLOS QoS levels

### 3.5. Handover

MCN-MR technologies could also help improve the cellular QoS experienced in handover (HO) areas. In cellular systems, a handover process is launched when the quality of a cellular connection with a given cell degrades, and other nearby cells can provide better quality levels. Certain standards (e.g. UMTS) allow for the possibility of MSs to be simultaneously connected to several BSs (soft handovers) to improve their QoS. However, HO areas usually correspond to cell edges with low signal levels, and not all standards support soft handovers (e.g. HSDPA).

Urban field trials have reproduced the scenario illustrated in Figure 7 in order to demonstrate how MCN-MR technologies could help improve the QoS in HO areas. In this scenario, a MS moves from one cell to a neighbouring cell in order to force a handover (in this case, Nemo Handy did not lock to a specific BS). Figure 7 represents the HO area as the area with cellular coverage from the two cells. It is important to note that HSDPA does not support soft HO, and therefore the MS is always connected to one BS. Figure 7 shows the *normalized throughput* measured in the case of a single-hop HSDPA cellular link as the MS moves from the source cell to the target cell. As shown in Figure 7, the performance significantly degrades (60-70% on average) in the HO area due to the lower signal level. The performance improves rapidly when the HO process ends and the MS connects to the target cell offering better signal quality.

MCN-MR technologies can help improving the QoS during a HO by replacing the single-hop cellular connection when the MS enters the HO area with an MCN-MR connection using a hybrid MN located outside the HO area and with good propagation conditions to the MS or destination node (for example, a hybrid MN located in the same street). This scenario was reproduced in Figure 7 using a hybrid MN that is 40 meters closer to the serving BS than the destination node. The *normalized throughput* as the MCN-MR destination node crosses the HO area is significantly higher than in the case of a single-hop cellular link. During the experiment, the hybrid MN did not enter the HO area, and the MCN-MR connection was capable of providing the destination node with QoS levels experienced outside the HO area. These unique experimental results show that MCN-MR technologies could significantly improve the QoS as mobile nodes cross HO areas.

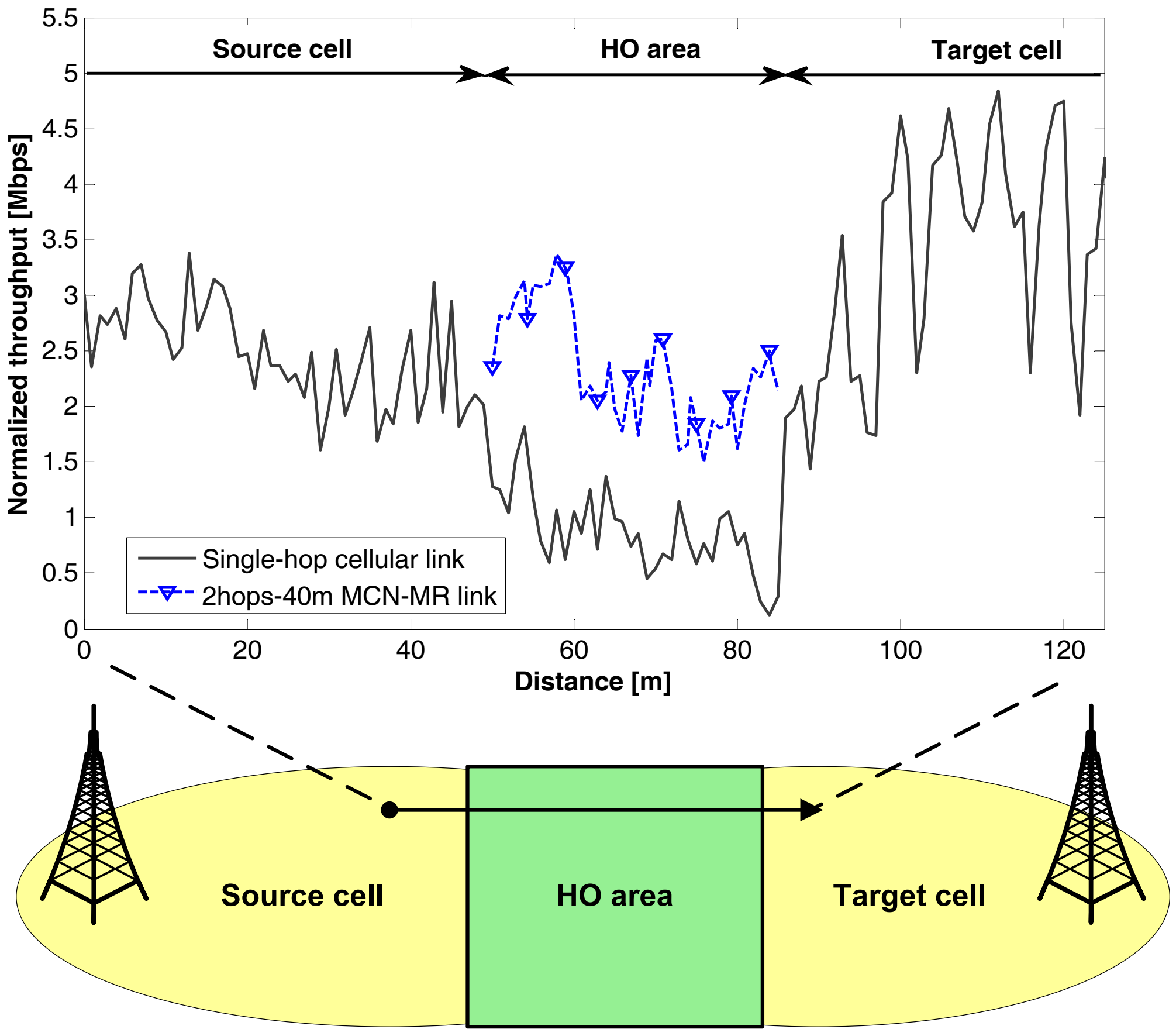

Figure 7. MCN-MR capacity to improve QoS in handover areas

**3.6. Energy Efficiency**

Previous studies (e.g. [8] and [9]) emphasized the possible energy consumption benefits of multi-hop cellular technologies. To experimentally evaluate this benefit, additional field tests have been conducted. These tests concentrate on uplink transmissions since energy gains for downlink transmissions are expected from the power management and saving techniques implemented at BSs. To measure such gains, access to operational BSs would have been necessary which was not possible for the authors at the time this study was conducted.

The field tests compare the energy consumed by a single-hop uplink HSUPA cellular transmission with that consumed by a 2-hop MCN-MR connection. In the case of the single-hop uplink HSUPA test, the MS was static and located inside a building (the MS was 30 meters away from the entrance of the building). The serving cellular BS was located 200 meters away from the entrance of the building. The BS had LOS conditions with the entrance of the building and NLOS conditions with the MS. In the case of the 2-hop MCN-MR connection, the MS gained access to the BS through a hybrid MN located at the entrance of the building and experiencing LOS conditions both with the BS and the MS. The energy consumption was measured using Nokia's Energy Profiler application. The measured energy consumption did not consider that corresponding to screen-backlight, Bluetooth and other applications/processes running in the background.

The conducted field tests showed that the single-hop uplink HSUPA cellular transmission from the MS to the BS required on average a transmission power of 15dBm and consumed 1.06µJ/bit. On the other hand, the cellular transmission from the hybrid MN to the BS required on average a transmission power of -20dBm and consumed 0.23µJ/bit. The 802.11g transmission from the MS to the hybrid MN consumed 0.29µJ/bit[5]. As a result, the MCN-MR connection only consumed 0.52µJ/bit, which represents a 50% reduction in the consumed energy compared to the single-hop uplink HSUPA cellular transmission.

**4. Conclusions**

The integration of cellular and ad-hoc technologies has been proposed to overcome certain limitations of conventional cellular architectures. Several analytical and simulation-based studies proved the benefits of such integration in terms of QoS, capacity and energy consumption among others. However, to date, no study had ever experimentally evaluated the QoS benefits of multi-hop cellular networks using mobile relays. In this context, this paper presents the results of an extensive field testing campaign conducted to evaluate the performance of MCN-MR networks, and the conditions under which they can overcome the limitations of conventional single-hop cellular systems. In particular, the conducted field tests have experimentally proved the capacity of MCN-MR technologies to extend the cellular radius, increase the QoS at large distances to the serving BS, in indoor environments, under NLOS propagation conditions, and while crossing handover areas. The article has also experimentally demonstrated the capacity of MCN-MR networks to reduce the energy consumption.

[5] The 802.11g transmission was performed using a Nokia N97 handset in order to measure the energy consumption using Nokia's Energy Profiler application

**ACKNOWLEDGMENTS**
This work has been supported by the Spanish Ministries of Science and Innovation, and Economy and Competitiveness, and FEDER funds under the projects TEC2008-06728 and TEC2011-26109, and by the Local Government of Valencia under the grant ACIF/2010/161. The authors would also like to acknowledge the support of Orange Spain.

**Javier Gozalvez** (j.gozalvez@umh.es) received an electronics engineering degree from the Engineering School ENSEIRB (Bordeaux, France), and a PhD in mobile communications from the University of Strathclyde, Glasgow, U.K. Since October 2002, he is with the Miguel Hernandez University of Elche, Spain, where he is currently an Associate Professor and Director of the UWICORE Laboratory. At UWICORE, he is leading research activities in the areas of wireless vehicular communications, radio resource management, heterogeneous wireless systems, and wireless system design and optimization. He currently serves as Mobile Radio

Senior Editor of IEEE Vehicular Technology Magazine, and previously served as AE of IEEE Communication Letters. He was TPC Co-Chair of the 2011 IEEE Vehicular Technology Conference-Fall, TPC Co-Chair of the 2009 IEEE Vehicular Technology Conference-Spring, and General Co-Chair of the 3rd ISWCS 2006. He is also the founder and General Co-Chair of the IEEE International Symposium on Wireless Vehicular communications (WiVeC) in its 2007, 2008, and 2010 editions. He has been elected to the Board of Governors of the IEEE Vehicular Technology Society (2011–2013), and to the IEEE Distinguished Lecturers program of the IEEE Vehicular Technology Society.

**Baldomero Coll-Perales** (bcoll@umh.es) received a Telecommunications Engineering degree from the Miguel Hernandez University (UMH) of Elche (Spain) in 2008. He received the Best Student awards in Telecommunications Engineering both by the UMH and the professional organization of Telecommunications Engineers. He then joined the UWICORE research laboratory to work on the development of networking and communication protocols for mobile relaying multi-hop cellular systems under the m-HOP and ICARUS research projects. He is currently pursuing his PhD studies with a fellowship from the Valencia Regional government. His research is focusing on multi-hop cellular networks, including networking, connectivity and resource management aspects.